\def \VersionAuthor {}
\ifdefined\VersionAuthor
	\newcommand{\AuthorVersion}[1]{#1}
	\newcommand{\FinalVersion}[1]{}
\else
	\newcommand{\AuthorVersion}[1]{}
	\newcommand{\FinalVersion}[1]{#1}
\fi

\documentclass[conference]{IEEEtran}
\IEEEoverridecommandlockouts

\usepackage[utf8]{inputenc}
\usepackage[english]{babel}

\usepackage[ruled,vlined,linesnumbered]{algorithm2e}
	\SetKwInOut{Input}{input}
	\SetKwInOut{Output}{output}

\usepackage{subcaption}

\usepackage{paralist} %

\newenvironment{ienumerate}
	{\ifdefined\VersionLong\begin{enumerate}\else\begin{inparaenum}[\itshape i\upshape)]\fi}
	{\ifdefined\VersionLong\end{enumerate}\else\end{inparaenum}\fi}

\newenvironment{oneenumerate}
	{\ifdefined\VersionLong\begin{enumerate}\else\begin{inparaenum}[1)]\fi}
	{\ifdefined\VersionLong\end{enumerate}\else\end{inparaenum}\fi}

\usepackage{amsmath,amssymb,amsfonts}

\ifdefined \VersionLong
	\newcommand{\LongVersion}[1]{\ifdefined\VersionWithComments{\color{red!40!black}\xspace#1\xspace}\else\xspace#1\xspace\fi}
	\newcommand{\ShortVersion}[1]{\ifdefined\VersionWithComments{\color{black!40}#1}\fi}
\else
	\newcommand{\LongVersion}[1]{\ifdefined\VersionWithComments{\color{black!40}\xspace#1\xspace}\fi}
	\newcommand{\ShortVersion}[1]{\ifdefined\VersionWithComments{\color{red!40!black}#1}\else#1\fi}
\fi

\ifdefined\VersionWithComments
	\usepackage{draftwatermark}
	\SetWatermarkText{draft}
	\SetWatermarkScale{3}
	\SetWatermarkColor[gray]{0.9}
\fi
\usepackage[svgnames,table]{xcolor}
\definecolor{darkblue}{rgb}{0, 0, 0.7}

\usepackage[
		pdfauthor={Étienne André and Aleksander Kryukov},%
		pdftitle={Parametric discrete non-interference in timed automata},
		breaklinks  = true,
		colorlinks  = true,
	\ifdefined \VersionWithComments
	\fi
		citecolor   = blue!50!blue,
		linkcolor   = darkblue,
		urlcolor    = blue!50!green,
	]{hyperref}

\ifdefined\VersionAuthor
	\usepackage[backend=biber,backref=true,style=alphabetic,url=false,doi=true,defernumbers=true,sorting=anyt,maxnames=99]{biblatex} %
	\renewbibmacro*{doi+eprint+url}{%
		\iftoggle{bbx:doi}
			{\color{black!40}\footnotesize\printfield{doi}}
			{}%
		\newunit\newblock
		\iftoggle{bbx:eprint}
			{\usebibmacro{eprint}}
			{}%
		\newunit\newblock
		\iftoggle{bbx:url}
			{\usebibmacro{url+urldate}}
			{}%
	}

\fi
\usepackage[capitalise,english,nameinlink]{cleveref} %
\crefname{line}{\text{line}}{\text{lines}} %

\newcommand{\defProblem}[3]
{%
\noindent\fcolorbox{black}{blue!15}{
	\begin{minipage}{.95\columnwidth}
		\textbf{#1 problem:}\\
		\textsc{Input}: #2\\
		\textsc{Problem}: #3
	\end{minipage}
}
	
	\smallskip
	
}

\usepackage{tikz}
\usetikzlibrary{arrows,automata}
\tikzstyle{every node}=[initial text=]
\tikzstyle{location}=[rectangle, rounded corners, minimum size=12pt, draw=black, fill=blue!10, inner sep=2pt]
\tikzstyle{invariant}=[draw=black, dotted, inner sep=1pt] %
\tikzstyle{final}=[double, fill=blue!50]
\tikzstyle{urgent}=[dotted, draw=red, very thick]

\tikzstyle{urgent}=[fill=yellow, thick, dotted] %
\tikzstyle{private}=[fill=red,thick]

\definecolor{coloract}{rgb}{0.50, 0.70, 0.30}
\definecolor{colorclock}{rgb}{0.4, 0.4, 1}
\definecolor{colordisc}{rgb}{1, 0, 1}
\definecolor{colorloc}{rgb}{0.4, 0.4, 0.65}
\definecolor{colorparam}{rgb}{1, 0.6, 0.0}

\definecolor{loccolor1}{rgb}{1, 0.3, 0.3}
\definecolor{loccolor2}{rgb}{0.3, 1, 0.3}
\definecolor{loccolor3}{rgb}{0.3, 0.3, 1}
\definecolor{loccolor4}{rgb}{1, 0.3, 1}
\definecolor{loccolor5}{rgb}{1, 1, 0.3}
\definecolor{loccolor6}{rgb}{0.3, 1, 1}
\definecolor{loccolor7}{rgb}{0.9, 0.6, 0.2}
\definecolor{loccolor8}{rgb}{0.7, 0.4, 1}
\definecolor{loccolor9}{rgb}{0.5, 1, 0.75}
\definecolor{loccolor10}{rgb}{0.8, 0.7, 0.6}
\definecolor{loccolor11}{rgb}{0.6, 0.7, 0.8}
\definecolor{loccolor12}{rgb}{0.2, 0.5, 0.9}
\definecolor{loccolor13}{rgb}{0.5, 0.9, 0.2}
\definecolor{loccolor14}{rgb}{0.9, 0.2, 0.5}
\definecolor{loccolor15}{rgb}{0.7, 0.7, 0.7}
\definecolor{loccolor16}{rgb}{0.8, 0.8, 0.5}

\newcommand{\styleact}[1]{\ensuremath{\textcolor{coloract}{{#1}}}}
\newcommand{\styleclock}[1]{\ensuremath{\textcolor{colorclock}{{#1}}}}
\newcommand{\styledisc}[1]{\ensuremath{\textcolor{colordisc}{\mathrm{#1}}}}
\newcommand{\styleloc}[1]{\ensuremath{\mathrm{#1}}}
\newcommand{\styleparam}[1]{\ensuremath{\textcolor{colorparam}{{#1}}}}
\usepackage{amsthm}
\theoremstyle{plain}

\newtheorem{proposition}{Proposition}

\theoremstyle{definition}
\newtheorem{definition}{Definition}
\newtheorem{example}{Example}

\theoremstyle{remark}
\newtheorem{remark}{Remark}

\ifdefined\VersionWithComments
	\usepackage[colorinlistoftodos,textsize=footnotesize]{todonotes}
\else
	\usepackage[disable]{todonotes}
\fi
\newcommand{\gennote}[3]{\todo[linecolor=#2,backgroundcolor=#2!25,bordercolor=#2]{#3: #1}\xspace}
\newcommand{\ea}[1]{\gennote{#1}{blue}{ÉA}}
\newcommand{\ak}[1]{{\gennote{#1}{purple}{AK}}}
\newcommand{\instructions}[1]{{\gennote{\bfseries #1}{red}{Instructions}}}
\newcommand{\reviewer}[2]{{\gennote{``#2''}{purple}{Reviewer #1}}}
\ifdefined \VersionWithComments
	\newcommand{\todoinline}[1]{\mbox{}{\color{red}{\textbf{TODO}\ifx#1\\\else:\ \fi #1}}} %
\else
	\newcommand{\todoinline}[1]{}
\fi

\usepackage{verbatim} %

\newcommand{\init}{_0}

\newcommand{\A}{\ensuremath{\mathcal{A}}}
\newcommand{\Actions}{\Sigma}
\newcommand{\ActionsH}{H}
\newcommand{\ActionsL}{L}
\newcommand{\action}{\ensuremath{a}}
\newcommand{\actionSilent}{\ensuremath{\epsilon}}
\newcommand{\ActionsIndices}{\zeta}
\newcommand{\assign}{\leftarrow}

\newcommand{\C}{C}

\newcommand{\Clock}{\mathbb{X}} %
\newcommand{\ClockCard}{H} %
\newcommand{\clock}{x} %
\newcommand{\clockx}{x} %
\newcommand{\clocky}{y} %
\newcommand{\clockinterf}{\clock_{\mathit{interf}}}
\newcommand{\clockval}{\mu} %
\newcommand{\ClocksZero}{\vec{0}}
\newcommand{\compOp}{\bowtie}

\newcommand{\edge}{e}
\newcommand{\Edges}{E}
\newcommand{\longuefleche}[1]{\stackrel{#1}{\longrightarrow}}
\newcommand{\longueflecheRel}[1]{\stackrel{#1}{\mapsto}}

\newcommand{\flecheRel}{{\rightarrow}}

\newcommand{\grandn}{{\mathbb N}}
\newcommand{\grandq}{{\mathbb Q}}
\newcommand{\grandqplus}{\grandq_{+}} %

\newcommand{\grandz}{{\mathbb Z}}
\newcommand{\guard}{g}

\newcommand{\InterfnH}{\ensuremath{\mathcal{I}\mathit{nterf}^n_{\ActionsH}}}
 
\newcommand{\invariant}{I}
\newcommand{\K}{K}

\newcommand{\loc}{\ensuremath{\ell}} %
\newcommand{\locinit}{\loc\init}
\newcommand{\Loc}{L} %
\newcommand{\Locations}{\mathsf{Loc}} %

\newcommand{\Param}{\mathbb{P}} %
\newcommand{\param}{p} %
\newcommand{\ParamCard}{M} %
\newcommand{\pval}{v} %
\newcommand{\R}{{\mathbb{R}}}
\newcommand{\Rgeqzero}{\R_{\geq 0}}

\newcommand{\sinit}{s\init} %
\newcommand{\somelocs}{T} %
\newcommand{\state}{\ensuremath{s}} %
\newcommand{\States}{S} %

\newcommand{\acc}{\ensuremath{\styleparam{\mathit{acc}}}}
\newcommand{\att}{\ensuremath{\styleact{\mathit{att}}}}
\newcommand{\ucs}{\ensuremath{\styleparam{\mathit{ucs}}}}
\newcommand{\varv}{\ensuremath{\styledisc{\mathtt{v}}}}

\newcommand{\hide}[2]{\ensuremath{#1_{\setminus #2}}}
\newcommand{\resets}{R}
\newcommand{\restrict}[2]{\ensuremath{#1_{|#2}}}

\newcommand{\reset}[2]{\ensuremath{[#1]_{#2}}}
\newcommand{\valuate}[2]{\ensuremath{#2(#1)}}
\newcommand{\stylealgo}[1]{\ensuremath{\textsf{#1}}}
\newcommand{\EFsynth}{\stylealgo{EFsynth}}

\newcommand{\imitator}{\textsf{IMITATOR}}
\newcommand{\uppaal}{\textsc{Uppaal}}

\ifdefined \VersionWithComments
 	\definecolor{colorok}{RGB}{80,80,150}
\else
	\definecolor{colorok}{RGB}{0,0,0}
\fi

\newcommand{\eg}{\textcolor{colorok}{e.\,g.,}\xspace}

\newcommand{\ie}{\textcolor{colorok}{i.\,e.,}\xspace}
\newcommand{\st}{\textcolor{colorok}{s.t.}\xspace}

\newcommand{\wrt}{\textcolor{colorok}{w.r.t.}\xspace}

\begin{document}

\title{Parametric non-interference in timed automata
\thanks{%
	\AuthorVersion{%
		This is the author %
			version of the manuscript of the same name published in the proceedings of the 25th International Conference on Engineering of Complex Computer Systems (\href{https://formal-analysis.com/iceccs/2020/}{ICECCS 2020}).
	}
	This work is partially supported by the ANR-NRF French-Singaporean research program \href{https://www.loria.science/ProMiS/}{ProMiS} (ANR-19-CE25-0015).}
}

\author{\IEEEauthorblockN{Étienne André}
\IEEEauthorblockA{
\textit{Université de Lorraine, CNRS, Inria, LORIA}\\
Nancy, France \\
\AuthorVersion{\protect\href{https://orcid.org/0000-0001-8473-9555}{\includegraphics[height=1em]{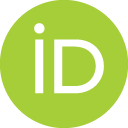}}}\FinalVersion{orcID:} \href{https://orcid.org/0000-0001-8473-9555}{0000-0001-8473-9555}}
\and
\IEEEauthorblockN{Aleksander Kryukov}
\IEEEauthorblockA{
\textit{Université de Lorraine, CNRS, Inria, LORIA}\\
Nancy, France}
}

\pagestyle{plain}

\maketitle

\begin{abstract}
	We consider a notion of non-interference for timed automata (TAs) that allows to quantify the frequency of an attack; that is, we infer values of the minimal time between two consecutive actions of the attacker, so that (s)he disturbs the set of reachable locations.
	We also synthesize valuations for the timing constants of the TA (seen as parameters) guaranteeing non-interference.
	We show that this can reduce to reachability synthesis in parametric timed automata.
	We apply our method to a model of the Fischer mutual exclusion protocol and obtain preliminary results.
\end{abstract}

\begin{IEEEkeywords}
security, non-interference, parametric timed automata
\end{IEEEkeywords}

\instructions{short papers should not exceed 6 pages, including figures, references, and appendices. Short paper submissions describe early-stage, ongoing or PhD research.}

\ea{hello}
\ak{hello}

\section{Introduction}

Timed automata (TAs) \cite{AD94} are a powerful formalism using which one can reason about complex systems involving time and concurrency.
Among various security aspects, \emph{non-interference} addresses the problem of deciding whether an intruder (or attacker) can disturb some aspects of the system.

In~\cite{BT03}, a decidable notion of non-interference is proposed to detect whether an intruder with a given frequency of the actions (s)he can perform is able or not to disturb the set of \emph{discrete} reachable behaviors (locations); that is, this notion can quantify the \emph{frequency} of an attack
In this paper,
we extend that definition in two different ways: first, by allowing some free parameters within the model---that becomes a parametric timed automaton (PTA)~\cite{AHV93}.
Second, by \emph{synthesizing} the admissible frequency for which the system remains secure, \ie{} for which the actions of the intruder cannot modify the set of reachable locations.

\paragraph{Contribution}
In this work, we propose a parametric notion of non-interference in timed automata that allows to \emph{quantify} the speed of the attacker necessary to disturb the model.
Our contribution is threefold:
\begin{oneenumerate}
	\item we define a notion of $n$-location-non-interference for timed automata;
	\item we show that checking this notion can reduce to reachability synthesis in PTAs;
	\item we model a benchmark from the literature, spot and correct an error in the original model, and we automatically infer using \imitator{}~\cite{AFKS12} parameter valuations for which the system is $n$-location-non-interfering.
\end{oneenumerate}

\paragraph{Related work}
It is well-known (see \eg{} \LongVersion{\cite{Kocher96,FS00,BB07,KPJJ13,BCLR15}}\ShortVersion{\cite{Kocher96,BCLR15}}) that time is a potential attack vector against secure systems.
That is, it is possible that a non-interferent (secure) system can become interferent (insecure) when timing constraints are added~\cite{GMR07}.
In~\cite{BDST02,BT03}, a first notion of \emph{timed} non-interference is proposed, based on traces and locations.
The latter is decidable as it reduces to the reachability problem for TAs~\cite{AD94}.
In~\cite{GMR07}, Gardey \emph{et al.}\ define timed strong non-deterministic non-interference (SNNI) based on timed language equivalence between the automaton with hidden low-level actions and the automaton with removed low-level actions. \LongVersion{Furthermore, they show that the problem of determining whether a timed automaton satisfies SNNI is undecidable. In contrast, timed cosimulation-based SNNI, timed bisimulation-based SNNI and timed state SNNI are decidable.}

In~\cite{Cassez09}, the problem of checking opacity for timed automata is considered: \LongVersion{even for the restricted class of event-recording automata~\cite{AFH99}, }it is undecidable whether a system is opaque, \ie{} whether an attacker can deduce whether some set of actions was performed, by only observing a given set of observable actions (with their timing).
In~\cite{AS19}, we proposed an alternative (and decidable) notion of opacity for timed automata, in which the intruder can only observe the \emph{execution time} of the system.
We also extend this notion to PTAs, and propose a procedure to automatically synthesize internal timings and admissible execution times for which the system remains opaque.

In~\cite{VNN18}, Vasilikos \emph{et al.}\ define the security of timed automata in term of information flow using a bisimulation relation and develop an algorithm for deriving a sound constraint for satisfying the information flow property locally based on relevant transitions.

In~\cite{BCLR15}, Benattar \emph{et al.}\ study the control synthesis problem of timed automata for SNNI. That is, given a timed automaton, they propose a method to automatically generate a (largest) sub-systems such that it is non-interferent if possible.
Different from the above-mentioned work, our work considers parametric timed automata, \ie{} timed systems with unknown design parameters, and focuses on synthesizing parameter valuations which guarantee non-interference.
In~\cite{NNV17}, the authors propose a type system dealing with non-determinism and (continuous) real-time, the adequacy of which is ensured using non-interference.
We share the common formalism of TA; however, we mainly focus on non-interference seen as the set of reachable locations, and we \emph{synthesize} internal parts of the system (clock guards), in contrast to~\cite{NNV17} where the system is fixed.

\paragraph{Outline}
In \cref{section:preliminaries}, we recall the necessary preliminaries, including non-interference for TAs.
In \cref{section:definition}, we define the problem of parametric location-non-interference for (P)TAs, and we draft a solution reducing to reachability synthesis.
In \cref{section:fischer}, we propose a new model for the Fischer mutual exclusion protocol.
In \cref{section:experiments}, we apply \imitator{} to this model, and obtain preliminary results ensuring non-interference.
We sketch future directions of research in \cref{section:conclusion}.

\section{Preliminaries}\label{section:preliminaries}

\LongVersion{
\subsection{Clocks, parameters and guards}
}

We assume a set~$\Clock = \{ \clock_1, \dots, \clock_\ClockCard \} $ of \emph{clocks}, \ie{} real-valued variables that evolve at the same rate.
A clock valuation is\LongVersion{ a function}
$\clockval : \Clock \rightarrow \Rgeqzero$.
We write $\ClocksZero$ for the clock valuation assigning $0$ to all clocks.
Given $d \in \Rgeqzero$, $\clockval + d$ \ShortVersion{is}\LongVersion{denotes the valuation} \st{} $(\clockval + d)(\clock) = \clockval(\clock) + d$, for all $\clock \in \Clock$.
Given $\resets \subseteq \Clock$, we define the \emph{reset} of a valuation~$\clockval$, denoted by $\reset{\clockval}{\resets}$, as follows: $\reset{\clockval}{\resets}(\clock) = 0$ if $\clock \in \resets$, and $\reset{\clockval}{\resets}(\clock)=\clockval(\clock)$ otherwise.

We assume a set~$\Param = \{ \param_1, \dots, \param_\ParamCard \} $ of \emph{parameters}\LongVersion{, \ie{} unknown constants}.
A parameter {\em valuation} $\pval$ is\LongVersion{ a function}
$\pval : \Param \rightarrow \grandqplus$.
We assume ${\compOp} \in \{<, \leq, =, \geq, >\}$.
A guard~$\guard$ is a constraint over $\Clock \cup \Param$ defined by a conjunction of inequalities of the form
$\clock \compOp \sum_{1 \leq i \leq \ParamCard} \alpha_i \param_i + d$, with
	$\param_i \in \Param$,
	and
	$\alpha_i, d \in \grandz$.
Given~$\guard$, we write~$\clockval\models\pval(\guard)$ if %
the expression obtained by replacing each~$\clock$ with~$\clockval(\clock)$ and each~$\param$ with~$\pval(\param)$ in~$\guard$ evaluates to true.

\subsection{Parametric timed automata}

\LongVersion{
Parametric timed automata (PTA) extend timed automata with parameters within guards and invariants in place of integer constants~\cite{AHV93}.
}

\begin{definition}[PTA]\label{def:uPTA}
	A PTA $\A$ is a tuple \mbox{$\A = (\Actions, \Loc, \locinit, \Clock, \Param, \invariant, \Edges)$}, where:
	\begin{ienumerate}
		\item $\Actions$ is a finite set of actions,
		\item $\Loc$ is a finite set of locations,
		\item $\locinit \in \Loc$ is the initial location,
		\item $\Clock$ is a finite set of clocks,
		\item $\Param$ is a finite set of parameters,
		\item $\invariant$ is the invariant, assigning to every $\loc\in \Loc$ a guard $\invariant(\loc)$,
		\item $\Edges$ is a finite set of edges  $\edge = (\loc,\guard,\action,\resets,\loc')$
		where~$\loc,\loc'\in \Loc$ are the source and target locations, $\action \in \Actions$, $\resets\subseteq \Clock$ is a set of clocks to be reset, and $\guard$ is a guard.
	\end{ienumerate}
\end{definition}
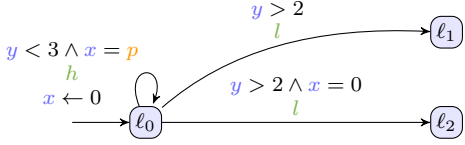
\begin{figure}[tb]
 
	\centering
	 \footnotesize

	\begin{tikzpicture}[scale=1, xscale=2, yscale=1.2, auto, ->, >=stealth']
 
		\node[location, initial] at (0, 0) (l0) {$\loc_0$};
 
		\node[location] at (2, 1) (l1) {$\loc_1$};
 
		\node[location] at (2, 0) (l2) {$\loc_2$};
 
		\path (l0) edge[loop above] node[left, align=center]{$\styleclock{y} < 3 \land \styleclock{x} = \styleparam{p}$\\$\styleact{h}$\\$\styleclock{x} \assign 0$} (l0);

		\path (l0) edge[bend left] node[above,align=center]{$\styleclock{y} > 2$\\$\styleact{l}$} (l1);

		\path (l0) edge[] node[above,align=center]{$\styleclock{y} > 2 \land \styleclock{x} = 0$\\$\styleact{l}$} (l2);
		
	\end{tikzpicture}
	\caption{A PTA example}
	\label{figure:example-PTA}

\end{figure}
\begin{example}
	Consider the PTA in \cref{figure:example-PTA}, containing two clocks~$\clockx$ and $\clocky$, and one parameter~$\param$.
	$\loc_0$ is the initial location.
	Observe that the transition to~$\loc_2$ can only be taken if the difference between $\clocky$ and $\clockx$ is larger than~2.
	This can only happen for selected valuations of the parameter~$\param$.
\end{example}

Given\LongVersion{ a parameter valuation}~$\pval$, we denote by $\valuate{\A}{\pval}$ the non-parametric structure where all occurrences of a parameter~$\param_i$ have been replaced by~$\pval(\param_i)$.
We denote as a \emph{timed automaton} any structure $\valuate{\A}{\pval}$\LongVersion{%
, by assuming a rescaling of the constants: by multiplying all constants in $\valuate{\A}{\pval}$ by the least common multiple of their denominators, we obtain an equivalent (integer-valued) TA\LongVersion{, as defined in~\cite{AD94}}}.

\LongVersion{
\subsubsection{Synchronized product of PTAs}
}

The \emph{synchronous product} (using strong broadcast, \ie{} synchronization on a given set of actions)\LongVersion{, or \emph{parallel composition},} of several PTAs gives a PTA.

\newcommand{\defSynchronizedProduct}{%

\begin{definition}[synchronized product of PTAs]\label{definition:parallel}
	Let $N \in \grandn$.
	Given a set of PTAs $\A_i = (\Actions_i, \Loc_i, (\locinit)_i, \Clock_i, \Param_i, \invariant_i, \Edges_i)$, $1 \leq i \leq N$,
	and a set of actions $\Actions_s$,
	the \emph{synchronized product} of $\A_i$, $1 \leq i \leq N$,
	denoted by $\A_1 \parallel_{\Actions_s} \A_2 \parallel_{\Actions_s} \cdots \parallel_{\Actions_s} \A_N$,
	is the tuple
		$(\Actions, \Loc, \locinit,  \Clock, \Param, \invariant, \Edges)$, where:
	\begin{enumerate}
		\item $\Actions = \bigcup_{i=1}^N\Actions_i$,
		\item $\Loc = \prod_{i=1}^N \Loc_i$,
		\LongVersion{\item} $\locinit = ((\locinit)_1, \dots, (\locinit)_N)$,
		\item $\Clock = \bigcup_{1 \leq i \leq N} \Clock_i$,
		\LongVersion{\item} $\Param = \bigcup_{1 \leq i \leq N} \Param_i$,
		\item $\invariant((\loc_1, \dots, \loc_N)) = \bigwedge_{i = 1}^{N} \invariant_i(\loc_i)$ for all $(\loc_1, \dots, \loc_N) \in \Loc$,
	\end{enumerate}
	and $\Edges{}$ is defined as follows.
	For all $\action \in \Actions$,
	let $\ActionsIndices_\action$ be the subset of indices $i \in 1, \dots, N$
	such that $\action \in \Actions_i$.
	For all  $\action \in \Actions$,
	for all $(\loc_1, \dots, \loc_N) \in \Loc$,
	for all \mbox{$(\loc_1', \dots, \loc_N') \in \Loc$},
	$\big((\loc_1, \dots, \loc_N), \guard, \action, \resets, (\loc'_1, \dots, \loc'_N)\big) \in \Edges$
	if:
	\begin{itemize}
		\item if $\action \in \Actions_s$, then
		\begin{oneenumerate}
			\item for all $i \in \ActionsIndices_\action$, there exist $\guard_i, \resets_i$ such that $(\loc_i, \guard_i, \action, \resets_i, \loc_i') \in \Edges_i$, $\guard = \bigwedge_{i \in \ActionsIndices_\action} \guard_i$, $\resets = \bigcup_{i \in \ActionsIndices_\action}\resets_i$, and,
			\item for all $i \not\in \ActionsIndices_\action$, $\loc_i' = \loc_i$.
		\end{oneenumerate}
		\item otherwise (if $\action \notin \Actions_s$), then there exists $i \in \ActionsIndices_\action$ such that
		\begin{oneenumerate}
			\item there exist $\guard_i, \resets_i$ such that $(\loc_i, \guard_i, \action, \resets_i, \loc_i') \in \Edges_i$, $\guard = \guard_i$, $\resets = \resets_i$, and,
			\item for all $j \neq i$, $\loc_j' = \loc_j$.
		\end{oneenumerate}
	\end{itemize}
\end{definition}

That is, synchronization is only performed on~$\Actions_s$, and other actions are interleaved.
\LongVersion{

}When $\Actions_s$ is not specified, it is assumed to be equal to the intersection of the sets of actions.
That is, given $\A_1$ over $\Actions_1$ and $\A_2$ over $\Actions_2$, $\A_1 \parallel \A_2$ denotes $\A_1 \parallel_{\Actions_s} \A_2$ where $\Actions_s = \Actions_1 \cap \Actions_2$.

}

\defSynchronizedProduct{}

\LongVersion{
\subsubsection{Concrete semantics of TAs}

Let us now recall the concrete semantics of TA.
}

\begin{definition}[Semantics of a TA]
	Given a PTA $\A = (\Actions, \Loc, \locinit, \Clock, \Param, \invariant, \Edges)$,
	and a parameter valuation~\(\pval\),
	the semantics of $\valuate{\A}{\pval}$ is given by the timed transition system (TTS) $(\States, \sinit, \flecheRel)$, with
	\begin{itemize}
		\item $\States = \{ (\loc, \clockval) \in \Loc \times \Rgeqzero^\ClockCard \mid \clockval \models \valuate{\invariant(\loc)}{\pval} \}$, %
		\LongVersion{\item} $\sinit = (\locinit, \ClocksZero) $,
		\item  $\flecheRel$ consists of the discrete and (continuous) delay transition relations:
		\begin{ienumerate}
			\item discrete transitions: $(\loc,\clockval) \longueflecheRel{\edge} (\loc',\clockval')$, %
				if $(\loc, \clockval) , (\loc',\clockval') \in \States$, and there exists $\edge = (\loc,\guard,\action,\resets,\loc') \in \Edges$, such that $\clockval'= \reset{\clockval}{\resets}$, and $\clockval\models\pval(\guard$).
			\item delay transitions: $(\loc,\clockval) \longueflecheRel{d} (\loc, \clockval+d)$, with $d \in \Rgeqzero$, if $\forall d' \in [0, d], (\loc, \clockval+d') \in \States$.
		\end{ienumerate}
	\end{itemize}
\end{definition}

    \LongVersion{Moreover we}\ShortVersion{We} write $(\loc, \clockval)\longuefleche{(\edge, d)} (\loc',\clockval')$ for a combination of a delay and discrete transition if
		$\exists  \clockval'' :  (\loc,\clockval) \longueflecheRel{d} (\loc,\clockval'') \longueflecheRel{\edge} (\loc',\clockval')$.

Given a TA~$\valuate{\A}{\pval}$ with concrete semantics $(\States, \sinit, \flecheRel)$, we refer to the states of~$\States$ as the \emph{concrete states} of~$\valuate{\A}{\pval}$.
A \emph{run} of~$\valuate{\A}{\pval}$ is an alternating sequence of concrete states of $\valuate{\A}{\pval}$ and pairs of edges and delays starting from the initial state $\sinit$ of the form
$\state_0, (\edge_0, d_0), \state_1, \cdots$
with
$i = 0, 1, \dots$, $\edge_i \in \Edges$, $d_i \in \Rgeqzero$ and
	$\state_i \longuefleche{(\edge_i, d_i)} \state_{i+1}$.
Given\LongVersion{ a state}~$\state=(\loc, \clockval)$, we say that $\state$ is reachable in~$\valuate{\A}{\pval}$ if $\state$ appears in a run of $\valuate{\A}{\pval}$.
By extension, we say that $\loc$ is reachable; and by extension again, given a set~$\somelocs$ of locations, we say that $\somelocs$ is reachable if there exists $\loc \in \somelocs$ such that $\loc$ is reachable in~$\valuate{\A}{\pval}$.
We denote by $\Locations(\valuate{\A}{\pval})$ the set of all locations reachable in~$\valuate{\A}{\pval}$.

\begin{example}
	Consider again the PTA~$\A$ in \cref{figure:example-PTA}.
	Let $\pval_1$ be such that $\pval_1(\param) = 1$.
	Then, $\loc_2$ is unreachable in $\valuate{\A}{\pval_1}$: at $\clockx = 1$, one can take a first time the self-loop over~$\loc_0$, yielding $\clocky = 1$ and $\clockx = 0$.
	The guard $\clocky > 2$ to~$\loc_2$ is not yet satisfied.
	Then at $\clockx = 1$, one can take a second time the self-loop over~$\loc_0$, yielding $\clocky = 2$ and $\clockx = 0$.
	The guard $\clocky > 2$ to~$\loc_2$ is still not satisfied.
	At $\clockx = 1$, the guard $\clocky < 3$ is not satisfied anymore, and the self-loop over~$\loc_0$ cannot be taken anymore.
	Therefore, $\Locations(\valuate{\A}{\pval_1}) = \{ \loc_0, \loc_1 \}$.
	
	Let $\pval_2$ be such that $\pval_2(\param) = 0.9$.
	This time, $\loc_2$ is reachable, by taking three times the self-loop over $\loc_0$ when $\clocky = 0.9$, $\clocky = 1.8$ and $\clocky = 2.7$ respectively.
	Therefore, $\Locations(\valuate{\A}{\pval_2}) = \{ \loc_0, \loc_1, \loc_2 \}$.
	
\end{example}
\subsection{Reachability synthesis}

We will use reachability synthesis to solve the problem in \cref{section:definition}.
This procedure, called \EFsynth{}, takes as input a PTA~$\A$ and a set of target locations~$\somelocs$, and attempts to synthesize all parameter valuations~$\pval$ for which~$\somelocs$ is reachable in~$\valuate{\A}{\pval}$.
$\EFsynth(\A, \somelocs)$ was formalized in \eg{} \cite{JLR15} and is a procedure that may not terminate, but that computes an exact result (sound and complete) if it terminates.
\EFsynth{} traverses the \emph{parametric zone graph} of~$\A$, which is a potentially infinite extension of the well-known zone graph of TAs (see, \eg{} \cite{ACEF09,JLR15}).
\begin{example}\label{example:EFsynth}
	Consider again the PTA~$\A$ in \cref{figure:example-PTA}.
	Let us compute the set of parameter valuations for which $\loc_2$ is reachable.
	$\EFsynth(\A, \{ \loc_2 \}) = 0 < \param < 1 \lor 1 < \param < 1.5 \lor 2 < \param < 3$.
	Intuitively, whenever $\param \in (0, 1)$, one can take multiple times the self-loop over~$\loc_0$ so that eventually the guard $\clocky > 2 \land \clockx = 0$ is satisfied;
	whenever $\param \in (1, 1.5)$, one can take exactly twice the self-loop over~$\loc_0$ so that the guard to~$\loc_2$ is satisfied;
	whenever $\param \in (2, 3)$, one takes a single time the self-loop over~$\loc_0$, and then the guard to~$\loc_2$ becomes satisfied.
	For other valuations, there is no way to reach~$\loc_2$.
\end{example}
\begin{remark}
	\EFsynth{} can also be used to compute \emph{unreachability} (or \emph{safety}) synthesis, by taking the \emph{negation} (\ie{} the complement of the valuations set) of the result.
\end{remark}
\begin{example}
	Consider again the PTA~$\A$ in \cref{figure:example-PTA}.
	Let us compute the set of parameter valuations for which $\loc_2$ is unreachable.
	$\neg \EFsynth(\A, \{ \loc_2 \})$ is $ \param = 0 \lor \param = 1 \lor \param \in [1.5, 2] \lor \param \geq 3$.
\end{example}
\subsection{Non-interference}

Often, non-interference is defined using a set of low-level actions and a set of high-level ones.
The idea is that an intruder is allowed to perform some high-level actions.
The non-interference property is satisfied whenever the system behavior in absence of high level actions is equivalent to its behavior, observed on low level actions, when high level actions occur~\cite{BT03}.

In the following, we assume a set of low-level actions $\ActionsL$ and a set of high-level actions $\ActionsH$.

\begin{definition}[restriction]
	Let $\A = (\Actions, \Loc, \locinit, \Clock, \Param, \invariant, \Edges)$ be a PTA with $\Actions = \ActionsL \uplus \ActionsH$ ($\uplus$ denotes disjoint union), and $\pval$ be a parameter valuation.
	The \emph{restriction of~$\valuate{\A}{\pval}$ to low-level actions}, denoted by~$\restrict{\valuate{\A}{\pval}}{\ActionsL}$,
	is defined as the automaton identical to $\valuate{\A}{\pval}$ except that any edge of the form $(\loc,\guard,\action,\resets,\loc')$ with $\action \in \ActionsH$ is discarded.
\end{definition}

\begin{definition}[hiding]
	Let $\A = (\Actions, \Loc, \locinit, \Clock, \Param, \invariant, \Edges)$ be a PTA with $\Actions = \ActionsL \uplus \ActionsH$, and $\pval$ be a parameter valuation.
	The \emph{hiding of high-level actions in~$\valuate{\A}{\pval}$}, denoted by~$\hide{\valuate{\A}{\pval}}{\ActionsH}$,
	is defined as the automaton identical to $\valuate{\A}{\pval}$ except that any edge of the form $(\loc,\guard,\action,\resets,\loc')$ with $\action \in \ActionsH$ is replaced with an edge $(\loc,\guard,\actionSilent,\resets,\loc')$.
\end{definition}

$\actionSilent$ is the special silent action.

In~\cite{BT03}, $n$-non-interference is defined as a concept ensuring that the low level behavior is unaffected by attacks which are separated by more than $n$ time units.
An attack is a high-level action decided by the attacker.
This concept helps to quantify the necessary \emph{attacking speed} of the attacker.

\begin{figure}[tb]
 
	\centering
	 \footnotesize

	\begin{tikzpicture}[scale=1, xscale=2, yscale=1.5, auto, ->, >=stealth']
 
		\node[location, initial] at (0, 0) (i0) {$\loc_0$};
 
		\node[location] at (2, 0) (i1) {$\loc_1$};
 
		\path (i0) edge[loop above] node[below left, align=center]{$\styleact{\ActionsL}$} (i0);
		
		\path (i1) edge[loop above] node[below left, align=center]{$\styleact{\ActionsL}$} (i1);
		
		\path (i0) edge node[above,align=center]{$\styleact{\ActionsH}$\\$\styleclock{\clockinterf} \assign 0$} (i1);
		
		\path (i1) edge[loop right] node[right,align=center]{$\styleclock{\clockinterf} \geq \styleparam{n}$\\$\styleact{\ActionsH}$\\$\styleclock{\clockinterf} \assign 0$} (i1);
		
	\end{tikzpicture}
	\caption{PTA $\InterfnH$ \cite{BT03}}
	\label{figure:interfnH}

\end{figure}

Let us now recall from~\cite{BT03} the (P)TA $\InterfnH$ in \cref{figure:interfnH}, where $\clockinterf$ is a local clock only used in~$\InterfnH$, and where $\ActionsL$ (resp.\ $\ActionsH$) denotes any transition labeled with an action $\action \in \ActionsL$ (resp.\ $\action \in \ActionsH$).
The idea is that this PTA allows the execution of high-level actions only when they are separated by at least $n$ time units.
Note that, in our setting, $n$ can be a timing parameter.

We now recall the concept of \emph{location-non-interference} (called \emph{state-non-interference} in~\cite{BT03}) that checks whether the set of locations (discrete states) reachable in the original automaton is identical to the set of locations reachable in the hiding\footnote{%
	The hiding of $\ActionsH$ is not strictly speaking necessary in our setting since we are interested in the reachability of \emph{locations} but we keep it for sake of consistency with~\cite{BT03}.
} of~$\ActionsH$ in the product of the original automaton with $\InterfnH$.

\begin{definition}[$n$-location-non-interference]\label{definition:nlocnoninterf}
	Let $\A = (\Actions, \Loc, \locinit, \Clock, \Param, \invariant, \Edges)$ be a PTA with $\Actions = \ActionsL \uplus \ActionsH$, and $\pval$ be a parameter valuation.
	Let $n \in \grandqplus$.
	$\valuate{\A}{\pval}$ is \emph{$n$-location-non-interfering} if $\Locations\big(\restrict{\valuate{\A}{\pval}}{\ActionsL}\big)$ is equal to
	$\Locations\big(\hide{(\valuate{\A}{\pval} \parallel \InterfnH )}{\ActionsH}\big)$ projected onto the locations of~$\valuate{\A}{\pval}$.
\end{definition}

By ``projected on $\valuate{\A}{\pval}$'', we mean the set
$ \{ \loc \mid \exists \loc' : (\loc, \loc') \in \Locations\big((\valuate{\A}{\pval} \parallel \InterfnH )\big) \} $.
In \cref{definition:nlocnoninterf}, the system is $n$-location-non-interfering if an intruder with the ability to disturb the system at most every $n$ time units is not able to modify the set of reachable locations.
Since the set of reachable locations is computable for TAs~\cite{AD94}, $n$-location-non-interference (for a given~$n$) is decidable for TAs~\cite{BT03}.

\begin{example}
	Consider again the PTA~$\A$ in \cref{figure:example-PTA}.
	Assume $\ActionsL = \{ l \}$ and $\ActionsH = \{ h \}$.
	That is, the intruder can take the self-loop over~$\loc_0$.
	Let $\pval_3$ be such that $\pval_3(\param) = 1.1$.
	First, note that $\Locations\big(\restrict{\valuate{\A}{\pval_3}}{\ActionsL}\big) = \{ \loc_0, \loc_1 \}$ since the transition to~$\loc_2$ is syntactically removed, preventing $\clockx$ to be reset.
	
	Fix $n = 1$.
	The product of $\valuate{\A}{\pval_3}$ with $\InterfnH$ prevents the system to synchronize faster than every 1~time unit on~$h$: therefore, taking the self-loop labeled with~$h$ when $\clocky = 1.1$ and $\clocky = 2.2$ respectively is possible, enabling the transition to~$\loc_2$ at $\clocky = 2.2$.
	This gives that $\Locations\big(\hide{(\valuate{\A}{\pval_3} \parallel \InterfnH )}{\ActionsH}\big)$ projected onto the locations of~$\valuate{\A}{\pval_3}$ is $\{ \loc_0, \loc_1, \loc_2 \}$.
	Therefore, $\valuate{\A}{\pval_3}$ is not 1-location-non-interfering.
	
	Now fix $n = 2$.
	In that case, the self-loop can be taken when $\clocky = 1.1$, but not when $\clocky = 2.2$ because the condition $\clockinterf \geq 2$ is not satisfied (recall that $\clockinterf$ is reset in $\InterfnH$ on the first transition labeled with~$h$, and therefore we have $\clockinterf = 1.1$ when $\clocky = 2.2$).
	So $\loc_2$ is unreachable.
	Therefore, $\valuate{\A}{\pval_3}$ is 2-location-non-interfering.
\end{example}
\section{Parametric location-non-interference}\label{section:definition}

In this work, we aim at considering a broader problem: instead of asking whether the intruder with a predefined power can disturb the system, we ask what is the power the intruder needs to perform a successful attack?
More precisely, we aim at computing the speed of the intruder needed to successfuly disturb the system: that is, for what valuations of $n$ is the system (not) $n$-location-non-interfering?

In addition, our PTA model can contain free parameters too; so the parameter is not only $n$ but also the PTA parameters\LongVersion{, \ie{} used in the timing constraints of the PTA}.

\LongVersion{
\subsection{Definition}
}

\defProblem{$n$-location-non-interference synthesis}
	{A PTA~$\A$ with parameters~$\Param$, a parameter $n$}
	{Synthesize valuations~$\pval$ of $\Param$ and of~$n$ such that $\valuate{\A}{\pval}$ is $n$-location-non-interfering.}

\LongVersion{
\subsection{Reachability synthesis}
}

Since our problem is \emph{location}-based, we can solve it using reachability synthesis techniques for PTAs, more precisely using \EFsynth{}.
The core idea is to synthesize valuations of $\Param \cup \{ n \}$ such that the set of reachable locations remains identical in both
$\restrict{\valuate{\A}{\pval}}{\ActionsL}$
and
$\hide{(\valuate{\A}{\pval} \parallel \InterfnH )}{\ActionsH}$.
This therefore reduces to a reachability synthesis problem.

\reviewer{1 (ICECCS)}{Comment: It seems to me that a theorem or proposition should be there to state the correctness of the algorithm -- \EFsynth{} solve the $n$-location-non-interference synthesis problem. }

Note that, since reachability-emptiness (\ie{} the emptiness of the valuations set for which a given (set of) location(s) is reachable) is undecidable for PTAs~\cite{AHV93,Andre19STTT}, reachability synthesis algorithms are not guaranteed to terminate.
(We discuss approximations later on.)

\LongVersion{
\subsection{Application to an example}
}

\begin{example}
	Consider again the PTA~$\A$ in \cref{figure:example-PTA}.
	Assume $\ActionsL = \{ l \}$ and $\ActionsH = \{ h \}$.
	First observe that $\loc_1$ (and of course $\loc_0)$ can be reached in $\valuate{\A}{\pval}$ regardless of the value of~$\param$.
	Second, for all $\pval$, $\loc_2$ is unreachable in $\restrict{\valuate{\A}{\pval}}{\ActionsL}$ since the self-loop on $\loc_0$ is syntactically removed.
	Therefore, for all~$\pval$, $\Locations\big(\restrict{\valuate{\A}{\pval}}{\ActionsL}\big) = \{ \loc_0, \loc_1 \}$.
	
	As a consequence, $n$-location-non-interference synthesis for~$\A$ reduces to unreachability synthesis of valuations of~$n$ and~$\param$ for which $\loc_2$ is unreachable.
	
	The result of $\neg \EFsynth\big(\hide{(\A \parallel \InterfnH )}{\ActionsH} , \{ \loc_2 \}\big)$ is:
	
	\noindent\begin{tabular}{@{}l l}
	& $(0 < \param < 1 \land n > \param)$ \\
	$\lor$ & $(\param = 1 \land n \geq 0)$ \\
	$\lor$ & $(1 < \param < 1.5 \land n > \param)$ \\
	$\lor$ & $(1.5 \leq \param \leq 2 \land n \geq 0)$ \\
	$\lor$ & $(\param \geq 3 \land n \geq 0)$\\
	\end{tabular}

	That is, for any valuation of~$\param$ and~$n$ within this constraint, the system is $n$-location-non-interfering, \ie{} the intruder cannot impact the set of reachable locations.

	This result can be intuitively explained as follows:
	whenever $\param < 1$ (first disjunct), if the intruder can act strictly slower than every $\param$ time unit ($n > \param$), only one self-loop on~$\loc_0$ can be taken, and $\loc_2$ is unreachable, and therefore the system is $n$-location-non-interfering.
	Whenever $\param = 1$ (second disjunct) or $1.5 \leq \param \leq 2$ (4th disjunct) or $\param \geq 3$ (last disjunct), we saw in \cref{example:EFsynth} that $\loc_2$ is unreachable, regardless of the value of~$n$.
	Finally, whenever $1 < \param < 1.5$ (3rd disjunct), $\loc_2$ is reachable iff the intruder can act strictly slower than every $\param$ time units.

\end{example}
\section{Application to the Fischer protocol}\label{section:fischer}
\subsection{The Fischer mutual exclusion protocol}\label{ss:Fischer}

The protocol proposed by Michael Fischer ensures that two processes never use a critical resource (often denoted by critical section) at the same time.
The protocol is based on the speed of the processes.

We consider here the protocol as studied in~\cite{BT03}:
suppose that two processes $P_1$ and~$P_2$ running in parallel compete for the critical section.
Assume that atomic reads and writes are permitted to a shared variable \texttt{v} (called \texttt{x} in~\cite{BT03}).
Assume also that every access to the shared memory containing~$\varv$ takes $\acc$ units of time.
Each process~$i$ executes the following code:

\begin{verbatim}
repeat
    await v = 0
    v := i
    delay b
until v = i
v := 0
(Critical section)
\end{verbatim}

The idea is that process $P_i$ is allowed into the critical section only when $\varv = i$;
``\texttt{await v=0}'' waits until \texttt{v} becomes~0;
``\texttt{delay b}'' waits exactly $b$ time units, which is ensured by the process local clock.
An assignment takes (at most) $a$ time units.
The modified model of~\cite{BT03} also considers that the maximum time needed to execute the critical section after \LongVersion{the variable }\texttt{v} is checked is~$\ucs$.

In~\cite{BT03}, the protocol is modeled using a network~$\A$ of TAs made of
\begin{ienumerate}
	\item process $P_i$, for $i=1,2$,
	\item an intruder, that can take an $\att$ transition anytime, nondeterministically changing $\varv$ to 0, 1 or~2, and
	\item a ``serializer'' responsible to model the value of~$\varv$ according to the access and modification requests of~$P_1$ and~$P_2$; recall that these operations take $\acc$ time units.
\end{ienumerate}

The idea is as follows: if the intruder is fast enough, (s)he can successfully disturb the system, \ie{} send both processes into the critical section at the same time, thus violating the mutual exclusion.
On the contrary, if the intruder is slow, its nondeterministic modifications of $\varv$ will have no effect on the security.

The crux of~\cite{BT03} is that the model is $n$-location-non-interfering iff the previously defined TA~$\A$ cannot reach a location where both processes are in the critical section at the same time.
A sufficient condition over $n$ and~$b$ to ensure $n$-location-non-interference is manually inferred and proved.

Our goal is to automatically infer conditions over $n$, $a$, $b$, $\acc$ and $\ucs$ guaranteeing $n$-location-non-interference.

\subsection{A modified Fischer model}
\subsubsection{An issue in the existing modeling}

While modeling the TA model of~\cite{BT03} using a model checker, we spotted a modeling issue, that makes the model wrong.
The issue lies in the serializer: the idea of the serializer is to ensure that two consecutive access or modification requests to~$\varv$ are separated by at least~$\acc$ time units;
but, due to the absence of invariants and of ``as soon as possible'' concept in the TAs of~\cite{BT03}, there exist runs of the TA such that the access to~$\varv$ can last forever, even if no other process is competing.
Although this cannot happen in reality, the model checker always reports that both processes can end up in the critical sections, whatever the values of~$b$ and~$n$ are.

Note that this mistake does not impact the definitions nor the overall reasoning of the benchmark application of~\cite{BT03}, as the authors use a \emph{manual} reasoning to infer the condition over~$n$ and~$b$.

\subsubsection{Our new model}

The purpose of the serializer is both to encode the value of~$\varv$, and to maintain a ``queue'' of requests to read or write accesses to the memory, ensuring that any two consecutive access is separated by~$\acc$ time units.
We therefore entirely rewrote the serializer, also impacting the model of the processes.

\begin{figure}
 
	\centering
	 \footnotesize

	 \scalebox{.9}{
	\begin{tikzpicture}[scale=1, xscale=2, yscale=1.5, auto, ->, >=stealth']
 
		\node[location, initial] at (0, 0) (i0) {$\styleloc{idle}$};
 
		\node[location] at (1.2, 0) (i1) {$\styleloc{reqT}$};
		
		\node[location] at (2.4, 0) (i2) {$\styleloc{active}$};
		\node[invariant, above=of i2, yshift=-25] {$\styleclock{\clock_i} \leq \styleparam{a}$};
		
		\node[location] at (3.6, 0) (i3) {$\styleloc{reqU}$};
		
		\node[location] at (3.6, -1) (i4) {$\styleloc{check}$};
		\node[invariant, left=of i4, xshift=25] {$\styleclock{\clock_i} \leq \styleparam{b}$};
		
		\node[location] at (3.6, -2) (i5) {$\styleloc{reqNP}$};
		
		\node[location] at (2.4, -2) (i6) {$\styleloc{preA}$};
		
		\node[location] at (1.2, -2) (i7) {$\styleloc{reqA}$};
		
		\node[location] at (0, -2) (i8) {$\styleloc{access}$};
		
		\node[location] at (0, -1) (i9) {$\styleloc{CS}$};
		\node[invariant, right=of i9, xshift=-25] {$\styleclock{\clock_i} \leq \ucs$};

		\path (i0) edge node {$\styleact{req_i}$} (i1);
		
		\path (i1) edge[bend right=55] node[above] {$\varv \neq 0$, $\styleact{\mathit{error}_i}$, $\styleclock{\clock_i} \assign 0$} (i0);
		
		\path (i1) edge node[below] {$\varv = 0$, $\styleclock{\clock_i} \assign 0$} node[above] {$\styleact{\mathit{try}_i}$} (i2);
		
		\path (i2) edge node[above] {$\styleclock{\clock_i} < \styleparam{a}$} node[below] {$\styleact{\mathit{req}_i}$} (i3);
		
		\path (i3) edge node[pos=0.3, right] {$\varv \assign i$,} node[pos=0.6, right] {$\styleclock{\clock_i} \assign 0$} node[pos=0.6, left] {$\styleact{\mathit{update}_i}$} (i4);
		
		\path (i4) edge node[pos=0.6, right] {$\styleclock{\clock_i}  = \styleparam{b}$} node[pos=0.4, left] {$\styleact{\mathit{req}_i}$} (i5);
		
		\path (i5) edge node[pos=0.5] {$\varv \neq i$} node[pos=0.4] {$\styleact{\mathit{no\_access}_i}$} (i0);
		
		\path (i5) edge node[above] {$\varv = i$} node[below] {$\styleact{\mathit{pre\_access}_i}$} (i6);
		
		\path (i6) edge node {$\styleact{req_i}$} (i7);
		
		\path (i7) edge node[below] {$\varv \assign 0$, $\styleclock{\clock_i} \assign 0$} node[above] {$\styleact{\mathit{access}_i}$} (i8);
		
		\path (i8) edge node[left] {$\styleclock{\clock_i} = 0$} (i9);
		
		\path (i9) edge node[align=center,pos=0.5, left] {$\styleclock{\clock_i} > 0$ \\ $\land \styleclock{\clock_i} \leq \ucs$} (i0);
		
	\end{tikzpicture}
	}
	
	\caption{Modified model of process $P_i$ for $i=1,2$}
	\label{figure:newProcessor}

\end{figure}
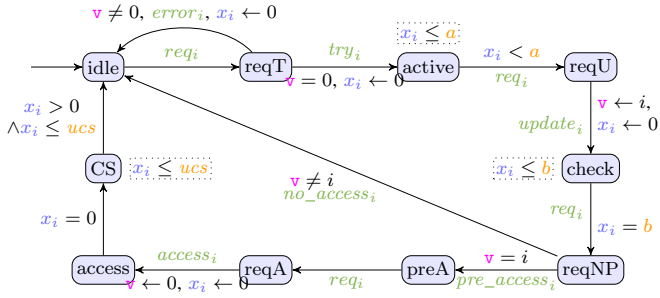

\paragraph{Processes}
We first give the modified model of process~$P_i$ in \cref{figure:newProcessor}.
Process~$P_i$ features a local clock~$\clock_i$ and can read or write~$\varv$.
The main modification \wrt{} to \cite[Fig.~6]{BT03} is the duplication of all locations: indeed, instead of performing a single action (\eg{} $\styleact{\mathit{try}_i}$) reading or writing~$\varv$, the process first performs a \emph{request} $\styleact{\mathit{req}_i}$, then followed by the action~$\styleact{\mathit{try}_i}$; the serializer is responsible for answering the request as soon as possible, but not earlier than~$\acc$ since the latest read or write action.
Then, the PTA follows the program given in \cref{ss:Fischer}:
$P_i$ first waits until $\varv = 0$, then updates it to~$i$;
then, it waits exactly $\styleparam{b}$ time units, and checks whether $\varv$ is still~$i$; if not, it moves back to the original location.
If $\varv = i$, the process sets $\varv$ to~0, enters the critical section and, after at most $\ucs$ time units, leaves it to go back to the \styleloc{idle} location.

\begin{figure}
 
	\centering
	\footnotesize

	 \scalebox{1}{
	\begin{tikzpicture}[scale=1, xscale=2, yscale=1.5, auto, ->, >=stealth']
 
		\node[location, initial] at (0, 0) (i0) {$\styleloc{idle}$};
 
		\path (i0) edge[loop above] node[left, xshift=-5, align=center] {$\styleact{\att}$ \\ $\varv \assign 0$} (i0);
		
		\path (i0) edge[loop right] node[align=center] {$\styleact{\att}$ \\ $\varv \assign 1$} (i0);
		
		\path (i0) edge[loop below] node[above left, xshift=-5, align=center] {$\styleact{\att}$ \\ $\varv \assign 2$} (i0);

	\end{tikzpicture}
	}
	
	\caption{Model of the intruder}
	\label{figure:intruder}

\end{figure}
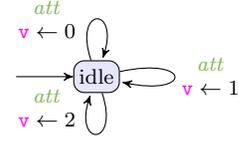

\paragraph{Intruder}
The intruder is almost identical to the one in~\cite{BT03}: it is a one-location PTA with three self-loops
synchronizing on~$\att$ and
setting $\varv$ to 0, 1 or~2 respectively (given in \cref{figure:intruder}).
Setting $\ActionsH = \{ \att \}$, the synchronization of the intruder with $\InterfnH$ ensures that the intruder may modify~$\varv$ with at least $\styleparam{n}$ time units since its last modification.

\paragraph{Serializer}
The seralizer automaton is both very simple and quite complex.
Its overall goal is very simple, and is to perform the following behavior: ``whenever a process ($P_1$, $P_2$ or the intruder) requests a write or read access to the memory, the serializer shall grant it as soon as possible, and $\acc$ time units later than the previous access---unless another process is also requesting access, in which case the request should be queued''.
Implementing this in a purely automata-based formalism is however cumbersome, much more than the simplistic serializer of~\cite{BT03}, that does not encompass for the ``as soon as possible'' concept.
This results in a PTA made of 12 locations and 91 transitions.\footnote{%
	All models and results are available at \href{https://www.imitator.fr/static/ICECCS20}{\nolinkurl{www.imitator.fr/static/ICECCS20}}.
}

\section{Experiments}\label{section:experiments}

We model the aforementioned automata using the \imitator{}~\cite{AFKS12} parametric timed model checker (version 2.12 ``Butter Lobster'').
We then run safety synthesis (\ie{} the negation of \EFsynth{}, implemented in \imitator{}) so as to synthesize parameter valuations for which mutual exclusion is guaranteed, \ie{} both processes cannot be in the \styleloc{CS} location at the same time.

\subsection{An approximated result}
Running \imitator{} on the model, the analysis does not terminate; this is not surprising as \EFsynth{} is not guaranteed to terminate due to the undecidability of the reachability-emptiness for PTAs~\cite{AHV93}.
A closer look at the analysis let us realize that, after passing the depth of 24 (\imitator{} explores the zone graph in a breadth-first search manner), no new constraint is synthesized, until at least a depth of 1000 (after which we interrupted the analysis, after about 20~hours of processing).
The resulting constraint (that does not change after depth~24, reached in about 650~s) is made of 22 disjunctions of convex constraints over the system parameters.

A property of \EFsynth{} is that it returns an under-approximation of the constraint when interrupted; when its negation (safety synthesis) is run, an \emph{over-approximation} is returned.
That is to say, the obtained result contains all possible valuations for which non-interference is satisfied, but the result may also potentially contain valuations for which the system is \emph{not} non-interfering.
We therefore \emph{tested} valuations from our constraint using the non-parametric timed model checker \uppaal{}~\cite{LPY97}.
We randomly picked up several dozens of parameter valuations, and checked using \uppaal{} that non-interference is satisfied iff the valuation belongs to the resulting constraint.
The \uppaal{} model is identical to the \imitator{} model, but is non-parametric, and therefore the analysis is guaranteed to terminate (depending on the parameter valuations, termination is obtained within a few seconds).
This does not formally prove that our result is exact (sound and complete), but increases the degree of confidence.
Proving the exactness of our constraint (or developing a new synthesis algorithm able to detect that the constraint is exact and to terminate the analysis) is among our future work.

\subsection{Interpretation}

The 22 disjunctions of convex constraints give a set of conditions for which the system is non-interfering, that is the mutual exclusion is guaranteed \emph{and} an attacker able to disturb the system at most every $n$ time units cannot succeed in violating mutual exclusion by its actions.

For sake of exemplification,
let us consider the first disjunct (the full constraint is available online):

\noindent\begin{tabular}{@{} l l}
	& $\styleparam{n} \geq 0$ \\
	$\land$ & $\styleparam{b} \geq \styleparam{\acc} + \styleparam{n}$\\
	$\land$ & $\styleparam{b} \geq 3 \times \styleparam{\acc}$\\
	$\land$ & $\styleparam{a} > 0$\\
	$\land$ & $\styleparam{\acc} > \styleparam{\ucs} > 0$\\
\end{tabular}

Recall that $b$ is the waiting time before testing again the value of~$\varv$, $n$ is the minimum time between any two consecutive high-level actions (of the intruder),
$\acc$ is the memory access time,
and $\ucs$ is an upper bound on the time during which a process remains in the critical section.
This constraint ensures that mutual exclusion is guaranteed even when an attacker can change the value of~$\varv$ no faster than every~$n$ time units if the following conditions are satisfied:
\begin{itemize}
	\item the delay ($b$) is longer than the access time ($\acc$) and the minimum disturbance time ($n$); that is to say, even when the intruder modifies the system, the process can still detect it as its delay is long enough; this helps guaranteeing non-interference;
	\item the delay is longer than three access times ($3 \times \acc$); that is to say, the delay is long enough to detect whether the other process performs $\styleact{\mathit{try}_i}$, $\styleact{\mathit{update}_i}$ and $\styleact{\mathit{pre\_access}_i}$ during the delay; this helps guaranteeing validity of the mutual exclusion;
	\item and the memory access time is longer than the time during which a process remains in the critical section.
\end{itemize}

Further sufficient conditions ensuring non-interference are guaranteed by the other disjuncts (see Web page).

\section{Conclusion}\label{section:conclusion}

\paragraph{Conclusion}
We introduced a definition of $n$-location-non-interference, that aims at quantifying the necessary attacker frequency to be able to modify the set of reachable locations in a timed automaton.
Using the \imitator{} parametric timed model checker, we obtained preliminary results on an improved version of the Fischer mutual exclusion protocol.

\paragraph{Future works}
As we only obtained an over-approximation of the result, our first future work is to prove the exactness (soundness and completeness) of the obtained constraint, either by proving it using an \emph{ad-hoc} reasoning for our case study, or by developing new automated techniques allowing \imitator{} to terminate as soon as the constraint is indeed complete.
Alternatively, designing approximated techniques is another interesting direction.

While the general emptiness problem (the emptiness of the set of both timing parameter valuations \emph{and} admissible values of~$n$ for which the system is $n$-location-non-interfering) is very likely to be undecidable (due to the undecidability of reachability emptiness in~\cite{AHV93}), the more specific problem of deciding whether there exists a valuation of~$n$ for which the (non-parametric) system is $n$-location-non-interfering remains open.
Also, we aim at tackling efficient \emph{synthesis} of these sets, independently of the decidability issues.
More generally, proposing new state space reduction techniques dedicated to the problem of $n$-location-non-interference is among our future works.

	\newcommand{\CCIS}{Communications in Computer and Information Science}
	\newcommand{\ENTCS}{Electronic Notes in Theoretical Computer Science}
	\newcommand{\FAC}{Formal Aspects of Computing}
	\newcommand{\FI}{Fundamenta Informaticae}
	\newcommand{\FMSD}{Formal Methods in System Design}
	\newcommand{\IJFCS}{International Journal of Foundations of Computer Science}
	\newcommand{\IJSSE}{International Journal of Secure Software Engineering}
	\newcommand{\IPL}{Information Processing Letters}
	\newcommand{\JAIR}{Journal of Artificial Intelligence Research}
	\newcommand{\JLAP}{Journal of Logic and Algebraic Programming}
	\newcommand{\JLAMP}{Journal of Logical and Algebraic Methods in Programming} %
	\newcommand{\JLC}{Journal of Logic and Computation}
	\newcommand{\LMCS}{Logical Methods in Computer Science}
	\newcommand{\LNCS}{Lecture Notes in Computer Science}
	\newcommand{\RESS}{Reliability Engineering \& System Safety}
	\newcommand{\STTT}{International Journal on Software Tools for Technology Transfer}
	\newcommand{\TCS}{Theoretical Computer Science}
	\newcommand{\ToPNoC}{Transactions on Petri Nets and Other Models of Concurrency}
	\newcommand{\TSE}{{IEEE} Transactions on Software Engineering}

\ifdefined\VersionAuthor
	\renewcommand*{\bibfont}{\small}
	\printbibliography[title={References}]
\else
	\bibliographystyle{IEEEtran} %
	\bibliography{pnStateNonInter}
\fi

\LongVersion{

\newpage
\appendix

\subsection{Old}

\ea{here, please add the definition of the protocol (in the form of a program?)}

Basic version of Fischer's protocol is described in \cite{BT03}. We want to explain the extended version of this protocol from that article and show the \textit{serializer'}.

We have to processors $P_{1}$ and $P_{2}$ that are running in parallel. They are competing for a critical section. \textit{x} is a shared variable. Atomic reads and writes are allowed for \textit{x}. \textit{acc} time units is a time to access \textit{x}. Improved version of the algorithm for processes:\\ 
\texttt{repeat} 

\texttt{await x = 0;}

\texttt{x := i;}

\texttt{delay b} \\
\texttt{until x = i;} \\
\texttt{x := 0} \\
\texttt{Critical section;} 

Each process can be in critical section iff \texttt{x = i}. \texttt{await x = 0} waits until the value of x becomes 0. The statement \texttt{delay b} delays a process for \texttt{b} time units. \texttt{b} depends on the process clock. We assume that the assignment statement takes at most \textit{a} time units. The mutual exclusion is guaranteed iff a $<$ b.

TA for each process is presented in \cref{fig:protocolAutomata}. \textit{y} is a clock. Process need x = 0 to start the protocol p(to move to state 1). At state 1 it can assign x in time shorter than \textit{b}(by algorithm it should be \textit{a} instead of \textit{b} but we used \textit{b} to make TA a bit simpler). In other words, $(x := i)_{P_{i}}$ takes \textit{a} time units and a $<$ b. After, in state 2 process test \textit{x} for \textit{b} time units(\texttt{delay b}). If \textit{x} is still equal \textit{i} process set variable \textit{x} to 0 and goes to critical section, if \textit{x} is not equal \textit{i} it tries to do the algorithm again.

$(x=i)_{P_{i}}$ means a synchronization action with the serializer' (in \cref{fig:serializer}) that also represents actions by an intruder but we will talk about an intruder later. All processes are synchronized on value of \textit{x} but not on actions.

\begin{figure}
 
	\centering
	 \footnotesize

	\begin{tikzpicture}[scale=1, xscale=2, yscale=1.5, auto, ->, >=stealth']
 
		\node[location, initial] at (0, 0) (i0) {$x = 0$};
 
		\node[location] at (2, 0) (i1) {$x = 1$};
		
		\node[location] at (1, -2) (i2) {$x = 2$};
 
		\path (i0) edge[loop above] node{1, 2, 3} (i0);
		
		\path (i0) edge node[above,align=center]{4, 5} (i1);
		
		\path (i0) edge[bend right] node[left]{6, 7} (i2);
		
		\path (i1) edge[loop above] node{5, 8, 9} (i1);
		
		\path (i1) edge node[left]{6, 7} (i2);
		
		\path (i1) edge[bend left] node{1, 2} (i0);
		
		\path (i2) edge[loop right] node{7, 10, 11} (i2);
		
		\path (i2) edge node[right,align=center]{1, 2} (i0);
		
		\path (i2) edge[bend right] node[right,align=center]{4, 5} (i1);
		
	\end{tikzpicture}
	\caption{The serializer'}
	\label{figure:serializer}

\end{figure}

Designations for the serializer' model:
\begin{enumerate}
  \item y$\geq$acc, $(x\assign0)_{P_{i}}$, \{y\};
  \item y$\geq$acc, $(x\assign0)_{att}$, \{y\};
  \item y$\geq$acc, $(x=0)_{P_{i}}$, \{y\};
  \item y$\geq$acc, $(x\assign1)_{P_{1}}$, \{y\};
  \item y$\geq$acc, $(x\assign1)_{att}$, \{y\};
  \item y$\geq$acc, $(x\assign2)_{P_{2}}$, \{y\};
  \item y$\geq$acc, $(x\assign2)_{att}$, \{y\};
  \item y$\geq$acc, $(x=1)_{P_{1}}$, \{y\};
  \item y$\geq$acc, $(x\neq2)_{P_{2}}$, \{y\};
  \item y$\geq$acc, $(x=2)_{P_{2}}$, \{y\};
  \item y$\geq$acc, $(x\neq1)_{P_{1}}$, \{y\}.
\end{enumerate}

\subsection{Modeling Fischer using timed automata}

\ea{here, please give the overall modeling idea, and point out the mistake from \cite{BT03}}

Now let's see the automata PPS ($P_{1}$ $||$ $P_{2}$ $||$ serializer'). We introduced serializer' to ensure atomicity of the accesses and to control the value of \textit{x}: y $\geq$ \textit{acc} makes delay for processes between accesses to variable \textit{x} that unsures atomicity; states of the serializer allow to control the value of \textit{x}. For PPS automata att actions are not considered, thus transitions 2, 5, 7 ca be removed from this case. PPS can reach all possible state $\langle s_{1}, s_{2}, s  \rangle$ except $\langle critical, critical, s  \rangle$ because both processes cannot be in critical state at the same time by mutual exclusion.

\textit{ucs} is an upper bound. It's a time that needed to execute the \textit{critical section}. Authors of the article \cite{BT03} described the idea of this expression in the best way:
\begin{quote}
\textit{``The idea is that, because \textit{ucs $\leq$ a + b}, after executing successfully the protocol and before entering the critical section, a process can signal to the other that the protocol can be executed again. This is safe because the time for executing the critical section is less than or equal to the time taken by the protocol itself. Actually \textit{ucs}includes the time for assigning the value 0 to x."}
\end{quote}

Now we are going to see automata PPSI = ($P_{1}$ $||$ $P_{2}$ $||$ serializer' $||$ intruder). The intruder's actions belongs to high-level actions. The intruders automata is shown in \cref{fig:protocolIntruder}. 

\begin{figure}
 
	\centering
	 \footnotesize

	\begin{tikzpicture}[scale=1, xscale=2, yscale=1.5, auto, ->, >=stealth']
 
		\node[location, initial] at (0, 0) (i0) {$x = 0$};
 
		\node[location] at (2, 0) (i1) {$x = 1$};
 
		\path (i0) edge[bend left] node[above,align=center, pos=0.3]{true,$(x\assign0)_{att}$,\{y\}} (i1);
		
		\path (i0) edge node[above,align=center]{true,$(x\assign1)_{att}$,\{y\}} (i1);
		
		\path (i0) edge[bend right] node[below, pos=0.3]{true,$(x\assign2)_{att}$,\{y\}} (i1);
		
		\path (i1) edge[loop above] node{y$\geq$n,$(x\assign0)_{att}$,\{y\}} (i1);
		
		\path (i1) edge[loop right] node[above] {y$\geq$n,} node {$(x\assign1)_{att}$,} node[below] {\{y\}} (i1);
		
		\path (i1) edge[loop below] node{y$\geq$n,$(x\assign2)_{att}$,\{y\}} (i1);
		
	\end{tikzpicture}
	\caption{The intruder}
	\label{figure:protocolIntruder}

\end{figure}

We want to show that PPSI is $n$-state-non-interfering(\ak{link to the definition if necessary}), for some $n$. It means that we need to show that the set of reachable states of the system ($P_{1}$ $||$ $P_{2}$ $||$ serializer' $||$ intruder $||$ $Interf^n_{H}$) $\setminus$ H projected on the states of PSSI is equal to the set of reachable states of the ($P_{1}$ $||$ $P_{2}$ $||$ serializer' $||$ intruder $||$ $Inhib_{H}$). To satisfy the \ak{link to the definition if necessary} we just need to prove that the states $\langle critical, critical, s, 0 \rangle$ of ($P_{1}$ $||$ $P_{2}$ $||$ serializer' $||$ intruder $||$ $Interf^n_{H}$) $\setminus$ H is still unreachable.

\begin{figure}
\centering
\includegraphics[scale=0.3]{protocolAttack.jpg}
\caption{An attack to the protocol}
\label{fig:protocolAttack}
\end{figure}

Firstly, we show the case when the attack in \cref{fig:protocolAttack} is possible. For this goal we need to set $n$ to enough low value. For example, \textit{acc} = 1, \textit{b} = 6, \textit{ucs} = 6 and $n$ = 2. Each depicted interval corresponds to \textit{acc} time units.

By the first case we can formulate the proposition:

\begin{proposition}
For all  $n$ $>$ \textit{b} the system ($P_{1}$ $||$ $P_{2}$ $||$ serializer' $||$ intruder) is $n$-state-non-interfering.
\label{prop:NgraterB}
\end{proposition}

Now we set $n$ to value grater than \textit{b} (according to \cref{prop:NgraterB}) and we will get another result by the same attack. After intruder changed the value of \textit{x} for the first time ($(x := 1)_{att}$), he need to wait $n$ time units until next possibility to change \textit{x}. During $n$ time units $P_{1}$ will move to critical section and $P_{2}$ will move to state 0 (because it failed the testing in state 2) and will start the protocol again. Thus, $P_{1}$ and $P_{2}$ will never be in critical section at the same time.

After reading the article \cite{BT03} we checked proposed model for reachability(both processors in critical section at the same time) in IMITATOR and we've got that the explained solution has mistakes and mutual exclusion property can be violated.

We found several mistakes: 
\begin{enumerate}
    \item condition ucs $\leq$ a + b. We can assume the worst run of the protocol when $a$ has a big value but processor's clock valuation in transition from 1 to 2 could be very small (almost 0). Also for the worst case we assume that ucs = a + b. Thus, there is a case when the property can be violated simpler. And we can claim that this condition is wrong.  %
    \item ``as soon as possible'' (ASAP) concept. During protocol execution it can be a situation when, for example, the $1^{st}$ processor can stop at the $3^{rd}$ state and neither invariant nor serializer will force it to leave this state. During this stop the $2^{nd}$ processor can reach the same state. After, we can assume, that the $1^{st}$ processor enter the critial section and will stay there $ucs$ time units. Then, after $acc$ time units the $2^{nd}$ processor will do the same. In this case, the mutual exclusion property can be vilated if $ucs > acc$. Also both processors just can stop at the $3^{rd}$ state and don't move futher because nothing force them to do it. In the \cref{tab:asapExample} after comparison $x = 2$ there can be no more actions and process will not terminate.
    \item lack of invariants. We mentioned it before but it was just a part of a problem. There are no invaiants in the states 1, 2 and $critical$. Thus there can be an execution when processor can stuck in that states (deadlock) if it will stay there more than $a$, $b$ and $ucs$ time units respectively. Again, nothing forces processor to leave that states.
\end{enumerate}

\begin{table}[]
\begin{tabular}{|l|l|l|l|l|l|l|l|l|l|l|}
\hline
$P_1$       & x = 0 &       & x := 1    &           &           & x = 1 &           &       & ...   & $\infty$ \\ \hline
$P_2$       &       & x = 0 &           & x := 2    &           &       &           & x = 2 & ...   & $\infty$ \\ \hline
$Intr$  &       &       &           &           & x := 1    &       & x := 2    &       & ...   & $\infty$ \\ \hline
\end{tabular}
\caption{Lack of ASAP concept}
\label{tab:asapExample}
\end{table}

\section{Experiments}\label{section:experiments}
\subsection{Implementation}

We propose new processor, serializer and intruder models with changes that help fix the above problems:
\begin{enumerate}
    \item We added ``want'' states and action for each existing state in processor and intruder models.
    \item We added invariants in processor model.
    \item We changed a structure of serializer and added ``wait'' states.
    \item We added new transition in processor to avoid deadlock.
\end{enumerate}

\begin{figure}
 
	\centering
	 \footnotesize

	\begin{tikzpicture}[scale=1, xscale=2, yscale=1.5, auto, ->, >=stealth']
 
		\node[location, initial] at (0, 0) (i0) {$idle$};
 
		\node[location] at (1.2, 0) (i1) {$wantT$};
		
		\node[location, label=above:\textcolor{purple}{y1 $<$ a}] at (2.4, 0) (i2) {$active$};
		
		\node[location] at (3.6, 0) (i3) {$wantU$};
		
		\node[location, label=left:\textcolor{purple}{y1 $\leq$ b}] at (3.6, -1) (i4) {$check$};
		
		\node[location] at (3.6, -2) (i5) {$wantNP$};
		
		\node[location] at (2.4, -2) (i6) {$preA$};
		
		\node[location] at (1.2, -2) (i7) {$wantA$};
		
		\node[location] at (0, -2) (i8) {$access$};
		
		\node[location, label=right:\textcolor{purple}{y1 $\leq$ ucs}] at (0, -1) (i9) {$CS$};
 
		\path (i0) edge node {$iWant_1!$} (i1);
		
		\path (i1) edge[bend right=55] node[above] {x $\neq$ 0, $error_1!$, y1 $\assign$ 0} (i0);
		
		\path (i1) edge node[below] {x = 0, y1 $\assign$ 0} node[above] {$try_1!$} (i2);
		
		\path (i2) edge node[above] {y1 $<$ a} node[below] {$iWant_1!$} (i3);
		
		\path (i3) edge node[pos=0.3, right] {x $\assign$ 1,} node[pos=0.6, right] {y1 $\assign$ 0} node[pos=0.6, left] {$update_1!$} (i4);
		
		\path (i4) edge node[pos=0.6, right] {y1 = b} node[pos=0.4, left] {$iWant_1!$} (i5);
		
		\path (i5) edge node[pos=0.5] {x $\neq$ 1} node[pos=0.4] {$no\_access_1!$} (i0);
		
		\path (i5) edge node[above] {x = 1} node[below] {$pre_access_1!$} (i6);
		
		\path (i6) edge node {$iWant_1!$} (i7);
		
		\path (i7) edge node[below] {x $\assign$ 0, y1 $\assign$ 0} node[above] {$access_1!$} (i8);
		
		\path (i8) edge node[left] {y1 = 0} (i9);
		
		\path (i9) edge node[pos=0.5, left] {0 $<$ y1 \&} node[pos=0.2, left] {y1 $\leq$ ucs} (i0);
		
	\end{tikzpicture}
	\caption{New processor}
	\label{figure:newProcessor-old}

\end{figure}
\begin{figure}
 
	\centering
	 \footnotesize

	\begin{tikzpicture}[scale=1, xscale=2, yscale=1.5, auto, ->, >=stealth']
 
		\node[location, initial] at (0, 0) (i0) {$idle$};
 
		\node[location] at (1, 0) (i1) {$wantT$};
		
		\node[location] at (2.5, 0) (i2) {$active$};
		
		\node[location] at (4, 0) (i3) {$wantU$};
		
		\path (i0) edge node {$iWant_a!$} (i1);
		
		\path (i1) edge[bend right=70] node[below] {$att_2!$, x $\assign$ 2, yA $\assign$ 0} (i2);
		
		\path (i1) edge node[above] {x = 1, yA = 0} node[below] {$att_1!$} (i2);
		
		\path (i1) edge[bend left=80] node[above] {$att_0!$, x $\assign$ 0, yA $\assign$ 0}(i2);
		
		\path (i2) edge[bend right=20] node[below] {yA $\geq$ n, $iWant_a!$} (i3);
		
		\path (i3) edge[bend left=90] node {$att_2!$, x $\assign$ 2, yA $\assign$ 0} (i2);
		
		\path (i3) edge[bend right=20] node[above] {x $\assign$ 1, yA $\assign$ 0} node[below] {$att_1!$} (i2);
		
		\path (i3) edge[bend right=80] node[above] {$att_0!$, x $\assign$ 0, yA $\assign$ 0} (i2);
		
	\end{tikzpicture}
	\caption{New intruder}
	\label{figure:newIntruder}

\end{figure}

\begin{remark}
We have changed the name of states from the article \cite{BT03} for our new model of processor: 0 - idle, 1 - active, 2 - check, 3 - preA (preAccess), 4 - access, critical - CS.
\end{remark}

Firstly, we fixed lack of invariants caused deadlock that we mentioned before by adding invariants $y1 < a$, $y1 \leq b$ and $y1 \leq ucs$ in the states ``1'', ``2'', ``critical'' respectively. 

Secondly, we tried to apply ASAP concept and found a way by adding want/wait states in processor/intruder models and changing that structure of serializer. Now serializer have ``idle'', ``delay'', urgent and wait states. ``Idle'' is an initial state, from here these is no need to wait $acc$ time units to do action that is why after it we added states(for both processors and an intruder) and made them $urgent$. They have to be in the model because of new synchronized ``want'' actions but no time have to elapse there. After performing any action serializer will move to ``delay'' state. It has to stay here $acc$ time units and move to the ``Idle'' state or it can move to wait states earlier than $acc$ time units.

``Wait'' states work like a queue if $acc$ time units are not elapsed since last action. There are 7 wait states: ``w1'', ``w2'', ``wi'', ``w12'', ``w1i'', ``w2i'', ``w12i''. The name of each state show you what is in a queue. But we do not control the order of leaving these state,  because of it we can check more cases. The sense of the ``wait'' states is to force processor/intruder, if it is in a queue, to perform action after $acc$ times units and not later. 

For example, serializer entered the ``w1'' state and after the ``w12''. It means that $1^{st}$ processor requested for transition and after the $2^{nd}$ did the same but time since last action is less than $acc$. But it doesn't mean the $1^{st}$ processor have to perform action before the $2^{nd}$. Thus, during simulation we can check to cases: when serializer left ``w12'' state and move to the state ``w2'' ($1^{st}$ processor performed action first) or to the state ``w1''($2^{nd}$ processor performed action first).

As we mentioned before ``want'' actions are synchronized actions and were added to all models. ``Want'' states are added before all transitions that contain operations with shared variable $x$ in a processor and an intruder models. Thus, they have to wait somewhere until the valuation of serializer's clocks will not be the same as $acc$.

New transition ``error'' were added to the processor model to avoid deadlock when both processors in a ``wantT'' state and intruder set $x$ to 1 or 2.

\subsection{Synthesis results}

$\imitator$ is a primary application of this work because our goals is to derive parametric notion of non-interference and illustrate it with parametric version of Fischer's protocol. When we tried to simulate new model we faced a problem that $\imitator$ could not terminate (\texttt{./imitator Fischer1State.imi -mode EF -merge -incl}). Also we noticed that after a certain depth $\imitator$ didn't find any violating property cases. Simulation were stopped after around depth = 1000 and violating property cases did not appear after the depth = 24. After we limited depth search (\texttt{./imitator Fischer1State.imi -mode EF -merge -incl -depth-limit 100}) to get results for violating property cases that $\imitator$ found. We got disjunction of 22 constraints. For each constraint we derived several valuations and checked it in $\uppaal$. 

22 constraints and 1 or 3 valuation(s) for them. First five constraints are the same for both models. From 4 to 10 there are the constraints with only one valuation because of condition $acc > ucs$ that means you can not violate mutual exclusion property. For example, if we are in a case when both processors in a ``wantA'' state and the $1^{st}$ entered the ``CS'' state, it will leave it before the $2^{nd}$ will be able to enter ``CS'' state because of $acc > ucs$. All valuations were checked in \uppaal for both models, not only for the ``1 state model'' and in both models we got that, at least with our valuations, mutual exclusion property can not be violated. 

\subsection{Interpretation}

}

\end{document}